\documentclass[10pt,journal]{IEEEtran}

\usepackage{cite}
\usepackage{graphicx}
\usepackage{algorithm}
\usepackage{algorithmic}
\usepackage{url}
\usepackage{amsmath}
\usepackage{wrapfig}

\newtheorem{theorem}{Theorem}

\renewcommand{\algorithmicrequire}{\textbf{Initialization:}}
\renewcommand{\algorithmicensure}{\textbf{Iteration:}}

\begin{document}

\title{Delay Analysis of Hybrid WiFi-LiFi System}

\author{\authorblockN{Sihua Shao\authorrefmark{1}, Abdallah Khreishah\authorrefmark{1}, Moussa Ayyash\authorrefmark{2}}\\
\authorblockA{\authorrefmark{1}Department of Electrical and Computer Engineering, New Jersey Institute of Technology, Newark 07102, USA\\Email: ss2536@njit.edu, abdallah@njit.edu}\\\authorblockA{\authorrefmark{2}Department of Information Studies, Chicago State University, Chicago 60628, USA\\Email: mayyash@csu.edu}}



\twocolumn[
  \begin{@twocolumnfalse}
    \maketitle
    \begin{abstract}
      Heterogeneous wireless networks are capable of effectively leveraging different access technologies to provide a wide variety of coverage areas. In this paper, the coexistence of WiFi and visible light communication (VLC) is investigated as a paradigm. The delay of two configurations of such heterogeneous system has been evaluated. In the first configuration, the non-aggregated system, any request is either allocated to WiFi or VLC. While in the second configuration, the aggregated system, each request is split into two pieces, one is forwarded to WiFi and the other is forwarded to VLC. Under the assumptions of Poisson arrival process of requests and the exponential distribution of requests size, it is mathematically proved that the aggregated system provides lower minimum average system delay than that of the non-aggregated system. For the non-aggregated system, the optimal traffic allocation ratio is derived. For the aggregated system, an efficient solution for the splitting ratio is proposed. Empirical results show that the solution proposed here incurs a delay penalty (less than 3\%) over the optimal result.
    \end{abstract}
  \end{@twocolumnfalse}
]

\section{Introduction}
%

Heterogeneous wireless network, as a method to incorporate different access technologies, contains the potential capabilities of improving the efficiency of spectral resource utilization. The coexistence of WiFi (a local area wireless access technology) and visible light communication (VLC) (an emerging complementary wireless access technology), can be considered as a typical prototype system.

The demand for real-time Internet multimedia services (e.g. high-definition video) becomes prevalent these days. Although additional small cells and WiFi access points (APs) are deployed to accommodate the growing number of mobile users, the overwhelming demand for high-throughput and low-delay Internet services is rarely met. This increasing demand will result in decreasing the quality-of-service (QoS) - a ``spectrum crunch" \cite{kavehrad2013optical}. To alleviate this problem, novel approaches that utilize additional spectrum should be investigated.

VLC technology, also referred to as LiFi, considered as complementary means of communication to WiFi, has been extensively investigated by researchers in the recent few years \cite{wu2014visible}. As an excellent candidate for 5G wireless communication, VLC provides ultra wide bandwidth and efficient energy utilization. Weaknesses of VLC include short transmission range and vulnerability to obstacles. To make the best of the pros of both VLC and WiFi, heterogeneous networks incorporating these two techniques should be developed.

Many current research efforts have been paid towards developing heterogeneous networks incorporating both WiFi and VLC. A protocol, considering OFDMA, vertical handover (VHO) and horizontal handover (HHO) mechanisms for mobile terminals (MTs) to enable the mobility of users among different VLC APs and OFDMA system, is proposed in \cite{bao2014protocol}. The authors define a new metric, called spatial density, to evaluate the capacity of the heterogeneous network under the assumption of the Homogenous Poisson Point Process (HPPP) distribution of MTs. In \cite{hou2006vertical}, a fuzzy-logic (FL)-based decision-making algorithm is proposed for VHO in the hybrid radio and optical wireless system. The authors adopt the interruption duration of optical link as a criterion to determine the VHO between RF and optical links.

Regarding the bandwidth aggregation, a thorough survey of approaches in heterogeneous wireless networks has been presented in \cite{ramaboli2012bandwidth}. The challenges and open research issues in the design of bandwidth aggregation system, ranging from MAC layer to application layer, have been investigated in detail. This work also validates the feasibility of the heterogeneous WiFi-VLC system proposed here based on bandwidth aggregation. In \cite{guo2015parallel}, users connect to WiFi and VLC simultaneously. A parallel transmission MAC (PT-MAC) protocol containing CSMA/CA algorithm and the concept of parallel transmission is proposed. This protocol supports fairness among users in the hybrid VLC and WLAN network.



The above-mentioned works, which are primarily simulation-based studies, do not provide physical implementation of the WiFi-LiFi systems. In our previous work \cite{ShaoJOCN2015}, an aggregated WiFi-VLC system is presented and implemented using WiFi/VLC equipment and Linux Bonding driver. The realized WiFi-LiFi system aggregates a single WiFi link and a single VLC link, and provides improved throughput. This paper theoretically investigates {\it system delay} $D$, a critical QoS metric especially for multimedia applications \cite{rahaim2011hybrid}. Here, system delay is defined as the amount of the time from the arrival of the request until it is completely served.

In \cite{rahaim2011hybrid}, delay modeling of a hybrid WiFi-VLC system has been investigated. Each WiFi and VLC queue is observed as an M/D/1 queue, and the capacities with respect to the unstable delay points of WiFi only, asymmetric WiFi-VLC and hybrid WiFi-VLC systems are compared. An analytic model for evaluating the queueing delays and channel access times at nodes in wireless networks is presented in \cite{tickoo2008modeling}. The model provides closed form solutions for obtaining the values of the delay and queue length. This is done by modeling each node as a discrete time G/G/1 queue. However, these works do not investigate the delay modeling of a system with bandwidth aggregation.


This paper characterizes the downlink system delay of two WiFi-VLC heterogeneous systems. One of them is based on bandwidth aggregation and the other is not. The potential gain in terms of the minimum average system delay through aggregating the bandwidth of WiFi and VLC is also evaluated. The main contributions of this work include the following: (i) a generalized characterization of the system without bandwidth aggregation in terms of the optimal ratio of traffic allocation and the minimum average system delay; (ii) a near-optimal characterization of the minimum average system delay of the system that utilizes bandwidth aggregation; (iii) theoretical proof that verifies the lower minimum average system delay of the system based on bandwidth aggregation, when compared to that of the system without bandwidth aggregation; (iv) empirical evaluation of the benefits of the system that utilizes bandwidth aggregation over the system that does not. Note that it is possible to generalize the delay analysis of such heterogeneous system to other access technologies.

\section{System Model}\label{sec2}
\begin{figure}
\centering
\includegraphics[width=0.40\textwidth]{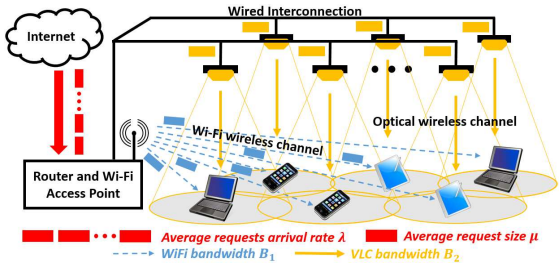}
  \caption{Heterogeneous system architecture}
  \label{fig_system_diagram}
\end{figure}

\begin{figure}
\centering
 \includegraphics[width=0.40\textwidth]{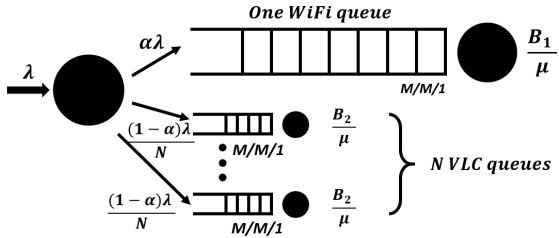}
  \caption{Queuing model representing the non-aggregated system model}
  \label{fig_non_aggregated}
\end{figure}
Fig.~\ref{fig_system_diagram} illustrates the heterogeneous system model. In the system model suggested, there is one WiFi AP and $N$ VLC APs. Each neighboring/interfering VLC AP uses different channel or different wavelength. Thus, the interference between VLC links is resolved. The VLC APs are assumed to cover the whole room, users can always connect to one of the VLC APs without outage problem. Additionally, the users' locations are assumed to be fixed and the handover delay triggered by movement is out of the scope of this paper. For a fixed-location scenario, the blockage issue of VLC will be negligible since no one would intentionally block the existing line-of-sight between the VLC APs and the clients. The traffic assigned to different VLC APs is evenly distributed. The process of requests arrival to the router is a Poisson process with rate $\lambda$. The size of each request is exponentially distributed with mean $\mu$. The downlink capacities of the WiFi and the VLC are $B_{1}$ and $B_{2}$, respectively, where $B_{1}<B_{2}$. In the non-aggregated system, any request is either allocated to the WiFi or the VLC. In the aggregated system, each request is split into two pieces. One of them is forwarded to the WiFi while the other is forwarded to one of the VLC APs. As a result, the system delay of each request is the maximum serving time of the two pieces. A new metric $\alpha$ is defined, to represent the traffic load assigned to WiFi and VLC. The next section describes in details the different configurations of two heterogeneous system models.

\section{System Delay Analysis}\label{sec3}
This section presents how to mathematically derive the minimum average system delay of the non-aggregated system. It provides a theoretical proof that the performance of the aggregated system is always better than that of the non-aggregated system in terms of the minimum average system delay. For the evaluation of the minimum average system delay of the aggregated system, an efficient solution is proposed to empirically incur a delay penalty (less than 3\%) over the optimal result and the comparison between the empirical results of the aggregated system and the delay performance of the non-aggregated system is also presented.

\subsection{The Non-aggregated System}
Let $\alpha$ denote the percentage of requests allocated to the WiFi. The non-aggregated system can be represented by the queuing model shown in Fig.~\ref{fig_non_aggregated}. Due to the assumption that requests are randomly forwarded to WiFi and VLC, the requests arrival to each queue is still a Poisson process. Requests arrive to WiFi and VLC queues with mean rates $\alpha\lambda$ and $(1-\alpha)\lambda/N$, respectively. The average serving rates of WiFi and VLC are $B_{1}/\mu$ and $B_{2}/\mu$, respectively.

\begin{theorem}
In the non-aggregated system model, the minimum average system delay is
\vspace{-0.1cm}
\begin{displaymath}
D_{min\_non\_agg}=\left\{ \begin{array}{rl}
&\frac{\mu N}{B_{2}N-\lambda\mu}, \mbox{ if $\frac{B_{2}N}{\lambda\mu}(1-\sqrt{\beta N})\geq1$}\\
&\frac{\lambda\mu(1+N)-B_{2}N(1-\sqrt{\beta N})^{2}}{\lambda[B_{2}N(\beta+1)-\lambda\mu]}, \mbox{ otherwise}
\end{array}\right.
\end{displaymath}
\end{theorem}

\begin{IEEEproof}
The optimization problem for minimizing the average system delay is formulated as follows:
\vspace{-0.1cm}
\begin{align}
\text{Objective:}~&\min~\alpha D_{WiFi}+(1-\alpha)D_{VLC}\nonumber\\
s.t.~&0\leq\alpha\leq1\nonumber\\
&\alpha\lambda<B_{1}/\mu\nonumber\\
&(1-\alpha)\lambda/N<B_{2}/\mu\nonumber
\end{align}

By calculating the derivative of the objective function and analyzing the candidate minimum points, the optimal $\alpha$ and the minimum objective value are obtained as follows:
\begin{align}
&\text{if}~\frac{B_{2}N}{\lambda\mu}(1-\sqrt{\beta N})\geq1,~\alpha_{opt}=0~\text{and}~D(0)=\frac{\mu N}{B_{2}N-\lambda\mu}\nonumber\\
&\text{otherwise},~\alpha_{opt}=\frac{\lambda\mu\sqrt{\beta}/(B_{2}N)+\sqrt{\beta}(\sqrt{\beta N}-1)}{\lambda\mu(\sqrt{\beta}+\sqrt{N})/(B_{2}N)}\nonumber\\
&\text{and}~D(\alpha_{opt})=\frac{\lambda\mu(1+N)-B_{2}N(1-\sqrt{\beta N})^{2}}{\lambda[B_{2}N(\beta+1)-\lambda\mu]}\nonumber
\end{align}
\end{IEEEproof}

\begin{figure}
\centering
 \includegraphics[width=0.40\textwidth]{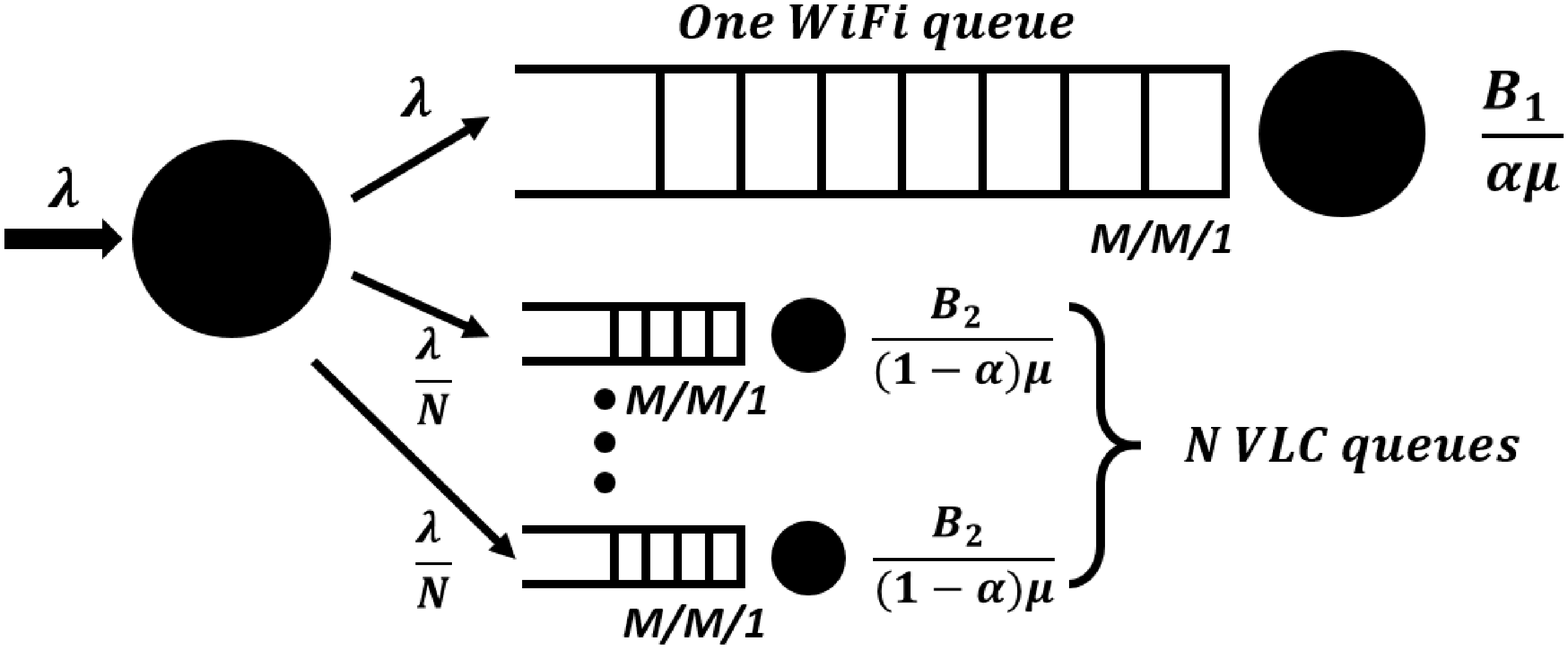}
  \caption{Queuing model representing the aggregated system model}
  \label{fig_aggregated}
\end{figure}

\subsection{The Aggregated System}

\begin{figure}
  \begin{center}
    \includegraphics[width=0.40\textwidth]{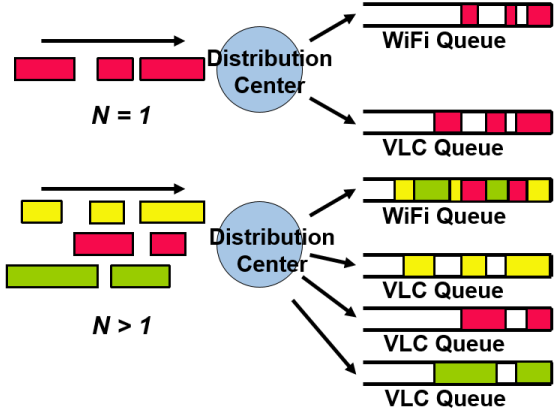}
  \end{center}
  \caption{Requests distribtuion in the aggregated system for two cases: $N = 1$ and $N > 1$}
  \label{fig_explain_alpha}
\end{figure}

\begin{figure*}
\centering
\includegraphics[width=\textwidth]{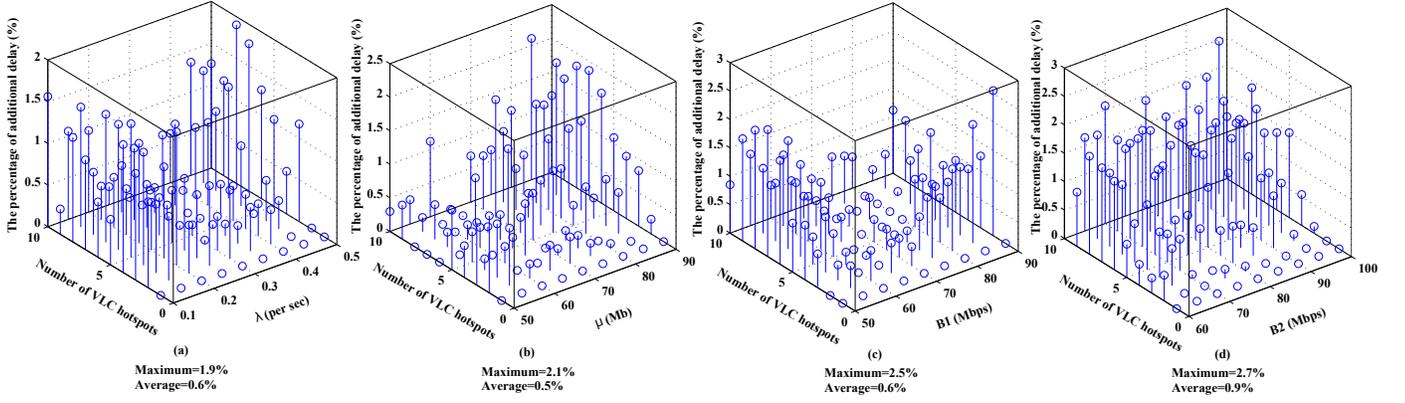}
\vspace{-10pt}
\caption{The percentages of additional delay caused by approximation in terms of (a) $\lambda$; (b) $\mu$; (c) $B_{1}$; (d) $B_{2}$, with $N$ varied from 1 to 10}
\label{fig_approximation}
\vspace{-8pt}
\end{figure*}

Let $\alpha$ denote the proportion of the size of each request that is allocated to the WiFi. The aggregated system can be represented by the queuing model shown in Fig.~\ref{fig_aggregated}. Assuming that the requests arrival are randomly and evenly distributed to each VLC queue, the requests arrival to each VLC queue is still a Poisson process. The average requests arrival rates for WiFi and VLC are $\lambda$ and $\lambda/N$. The average serving rates of WiFi and VLC are $B_{1}/(\alpha\mu)$ and $B_{2}/[(1-\alpha)\mu]$. The objective of the optimization problem can be expressed as minimizing $E[\max(D_{WiFi},D_{VLC})]$.

Fig.~\ref{fig_explain_alpha} represents the requests distribution to WiFi and VLC queues for two cases: $N=1$ and $N>1$. In Fig.~\ref{fig_explain_alpha}, it can be seen that when $N=1$, the delay of the VLC queue is fully correlated to that of the WiFi queue. Therefore, achieving the objective value of minimizing $E[\max(D_{WiFi},D_{VLC})]$ is equivalent to obtaining the optimal $\alpha$ from $E[D_{WiFi}]=E[D_{VLC}]$. However, when $N>1$, the WiFi queue contains different colored pieces of request, which are split from the requests flowing to different VLC APs. Each color represents a data stream destined to one VLC AP. The arrival times and the sizes of different colored pieces of request are independent while those of the same colored pieces of request are completely correlated. Specifically, due to the existence of red and green pieces of request (in Fig.~\ref{fig_explain_alpha}) in the WiFi queue, the departure times of the yellow pieces of request in the WiFi queue and the VLC queue are neither independent nor completely correlated. Hence, the complexity of computing the optimal alpha is severely exacerbated. Instead of searching for the optimal alpha by minimizing $E[\max(D_{WiFi},D_{VLC})]$, the objective is simplified as minimizing $\max(E[D_{WiFi}],E[D_{VLC}])$. For instance, let us assume that the delays of three pieces of request in WiFi are 1, 2 and 3 seconds respectively, and the delays of the corresponding three pieces of request in VLC are 2 seconds for all. As such, the objective value of $E[\max(D_{WiFi},D_{VLC})]$ will be 2.33 seconds while the objective value of $\max(E[D_{WiFi}],E[D_{VLC}])$ will be 2 seconds, which provides an underestimation of the traffic load. When the WiFi queue is overwhelmed, approximated $E[D_{WiFi}]$ will be lower than the real average request delay and vice versa. The error of this approximation approach depends on the congestion level of the WiFi queue. The error value has been further validated not to exceed 3\% by the results simulated in this paper. To determine the approximated value of the optimal $\alpha$ from the objective of minimizing $\max(E[D_{WiFi}],E[D_{VLC}])$, we make $E[D_{WiFi}]=E[D_{VLC}]$. Therefore, the approximated value of $\alpha$ is, $\alpha=(-b-\sqrt{b^{2}-4ac})/(2a)$, where $a=\lambda\mu(1-1/N)$, $b=-[B_{1}+B_{2}+\lambda\mu(1-1/N)]$, and $c=B_{1}$.

By simulating the aggregated system with the approximated $\alpha$, the percentages of additional delay caused by approximation are shown in Fig.~\ref{fig_approximation}. The values of the $\lambda, \mu, B_{1}, B_{2}$ are initially set as 0.5/s, 90 Mb, 50 Mpbs, 100 Mbps, respectively. In each plot, one of these four parameters is varied while keeping the other three fixed at the initial values. With $N$ varied from 1 to 10, it is noticed that the percentage of the maximum additional delay is 2.7\%, which is less than 3\%. Fig.~\ref{fig_approximation} (a)-(c), shows that, as $\lambda$, $\mu$ and $B_{1}$ increase, the percentage of the additional delay decreases initially and increases after reaching the minimum level. However, in Fig.~\ref{fig_approximation} (d), the percentage of the delay penalty does not change much. Since WiFi has the smaller bandwidth, maximum system delay of each request is more likely to be the system delay in WiFi than that in VLC. Therefore, the quantity of additional delay mainly depends on the level of congestion in WiFi queue. Fig.~\ref{fig_approximation} (a)-(c) shows that the percentage of additional delay has the minimum values  when $\lambda\approx0.33$, $\mu\approx58$ and $B_{1}\approx70$, respectively. When $\lambda<0.33$, $\mu<58$ and $B_{1}>70$, the approximation approach overestimates the congestion level of WiFi and causes additional traffic load allocated to VLC, and vice versa. Note that when $N=1$, the approximated solution proposed here will lead to the exact minimum average system delay of the aggregated system because the requests reached in each queue are fully correlated.

\begin{figure*}
\centering
\includegraphics[width=\textwidth]{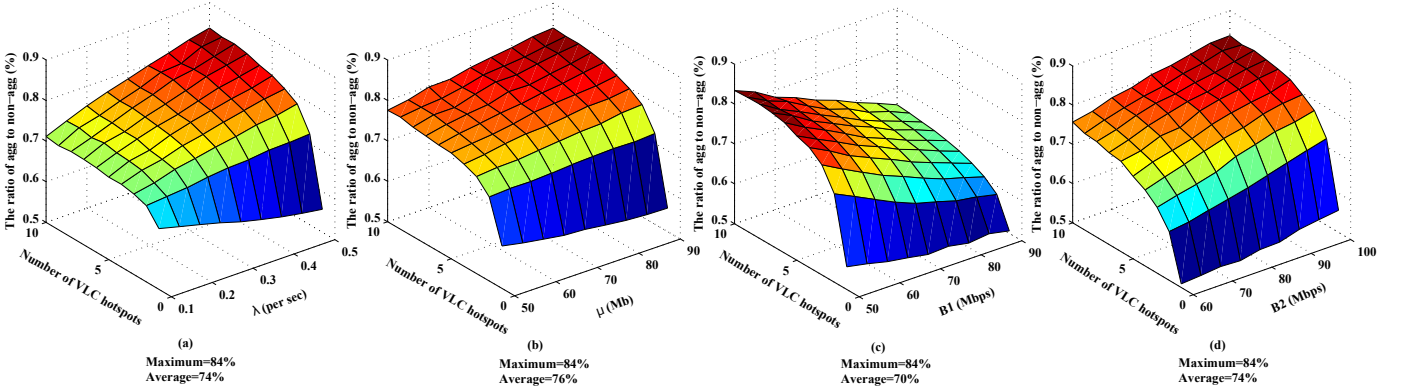}
\vspace{-10pt}
\caption{The ratio of the approximated minimum average system delay of the aggregated system to the minimum average system delay of the non-aggregated system in terms of (a) $\lambda$; (b) $\mu$; (c) $B_{1}$; (d) $B_{2}$, with $N$ varied from 1 to 10}
\label{fig_agg_vs_non}
\vspace{-8pt}
\end{figure*}

\subsection{Theoretical Analysis}
\begin{theorem}
The aggregated system has a lower minimum average system delay than that of the non-aggregated system.
\end{theorem}

\begin{IEEEproof}
The average system delays of the non-aggregated and the aggregated systems are
\begin{align}
&E[D_{non\_agg}]=\frac{\alpha}{B_{1}/\mu-\alpha\lambda}+\frac{1-\alpha}{B_{2}/\mu-(1-\alpha)\lambda/N}\nonumber\\
&E[D_{agg}]=E[\max(D_{WiFi},D_{VLC})]\nonumber\\
&=E[D_{WiFi}]+E[D_{VLC}]-E[\min(D_{WiFi},D_{VLC})]\nonumber
\end{align}
\vspace{-0.5cm}
\begin{align}
&\text{Note that},~E[D_{WiFi}]=\frac{1}{B_{1}/(\alpha\mu)-\lambda}=\frac{\alpha}{B_{1}/\mu-\alpha\lambda}\nonumber\\
&E[D_{VLC}]=\frac{1}{B_{2}/[(1-\alpha)\mu]-\lambda/N}=\frac{1-\alpha}{B_{2}/\mu-(1-\alpha)\lambda/N}\nonumber
\end{align}

Since $E[\min(D_{WiFi},D_{VLC})]$ is greater than zero, we have $E[D_{non\_agg}]>E[D_{agg}]$.

\end{IEEEproof}

\subsection{Empirical Analysis}
When applying the approximation method, the following question should be addressed: is the resulting minimum average system delay with approximated $\alpha$ of the aggregated system still lower than that of the non-aggregated system? To further investigate the comparison between the non-aggregated and the aggregated systems, the analytical results reached when applying the non-aggregated system are compared with the simulation results reached when applying the approximated aggregated system. The ratio of the approximated minimum average system delay of the aggregated system to the minimum average system delay of the non-aggregated system is used to demonstrate the practicability of the approximation approach. Fig.~\ref{fig_agg_vs_non} illustrates the comparison. The value settings of $\lambda, \mu, B_{1}, B_{2}$ and $N$ are the same as those in Fig.~\ref{fig_approximation}. As such, based on the simulation parameters, the approximated minimum average system delay of the aggregated system is at least 16\% lower than that of the non-aggregated system. The aggregation has diminishing gains over the non-aggregated system as the number of VLC APs increases and the ratio of WiFi bandwidth to VLC bandwidth decreases. This is due to the additional WiFi capacity which leads to decreasing the effect per VLC AP. Besides, the benefit of aggregating WiFi and VLC becomes less evident as $\lambda$ and $\mu$ are increasing. This is because increasing traffic load reduces the effect of efficient bandwidth utilization provided by aggregation.

\section{Conclusion}\label{sec4}
In this paper, two different configurations of a heterogeneous system are considered for aggregation and non-aggregation cases. Given the assumptions that requests arrive as a Poisson process and the request size is exponentially distributed, it is proved that the minimum average system delay of the aggregated system is lower than that of the non-aggregated system. An efficient method is proposed to approximate the optimal requests splitting ratio in the aggregated system. The analytical results when applying the non-aggregated system and simulation results when applying the aggregation system are also presented.


\bibliographystyle{IEEEtran}
\bibliography{Hybrid_System}

\end{document}